\documentclass{aa}  

\usepackage{graphicx}
\usepackage{txfonts}

\newcommand{\ms}{m\,s$^{-1}$}

\def\Teff{$T_{\mathrm{eff}}$}
\def\logg{\ensuremath{\log g}}
\def\vmic{$\upsilon_{\mathrm{mic}}$}

\def\vsini{\ensuremath{{\upsilon}\sin i}}
\def\kms{$\mathrm{km\,s}^{-1}$}
\def\ms{\hbox{\,m\,s$^{-1}$}}         %m.s -1
\def\m2s2{\hbox{\,m$^{2}$\,s$^{-2}$}} %m2.s -2
\def\kms{\hbox{\,km\,s$^{-1}$}}       %km.s -1
\def\vsini{\hbox{$v$\,sin\,$i$}}      %vsini
\def\Msun{\hbox{$\mathrm{M}_{\odot}$}}             %Msun
\def\Rsun{\hbox{$\mathrm{R}_{\odot}$}}
\def\Mjup{\hbox{$\mathrm{M}_{\rm Jup}$}}
\def\Rjup{\hbox{$\mathrm{R}_{\rm Jup}$}}
\begin{document} 

   \title{The GAPS Programme with HARPS-N at TNG}
   \subtitle{XIX. Atmospheric Rossiter-McLaughlin effect and improved parameters of KELT-9b\thanks{Based on observations made with the Italian Telescopio Nazionale Galileo (TNG) operated on the island of La Palma by the Fundacion Galileo Galilei of the INAF at the Spanish Observatorio Roque de los Muchachos of the IAC in the frame of the program Global Architecture of the Planetary Systems (GAPS).}
}
   \titlerunning{Atmospheric RML of KELT-9b}
   \authorrunning{F.~Borsa et al.}
 %  \subtitle{I. Overviewing the $\kappa$-mechanism}
    \author{F.~Borsa
          \inst{1},
          M.~Rainer\inst{2}, A.~S.~Bonomo\inst{3}, D.~Barbato\inst{4,3}, L.~Fossati\inst{5}, L.~Malavolta\inst{6}, V.~Nascimbeni\inst{7}, A.~F.~Lanza\inst{6}, 
          M.~Esposito\inst{8},
 L.~Affer\inst{9}, G.~Andreuzzi\inst{10,23}, S.~Benatti\inst{9}, K.~Biazzo\inst{6}, A.~Bignamini\inst{11}, M.~Brogi\inst{12,13,3}, I.~Carleo\inst{14}, R.~Claudi\inst{7}, R.~Cosentino\inst{10}, E.~Covino\inst{15}, M.~Damasso\inst{3}, S.~Desidera\inst{7}, A.~Garrido~Rubio\inst{9,16}, P.~Giacobbe\inst{3}, E.~Gonz{\'a}lez-{\'A}lvarez\inst{17}, A.~Harutyunyan\inst{10}, C.~Knapic\inst{11}, G. Leto\inst{6}, R.~Ligi\inst{1}, A.~Maggio\inst{9},  J.~Maldonado\inst{9}, L.~Mancini\inst{18,19,3}, A.~F.~M.~Fiorenzano\inst{10}, S.~Masiero\inst{24}, G.~Micela\inst{9}, E.~Molinari\inst{20}, I.~Pagano\inst{6}, M.~Pedani\inst{10}, G.~Piotto\inst{21}, L.~Pino\inst{22}, E.~Poretti\inst{1,10}, G.~Scandariato\inst{6}, R.~Smareglia\inst{11}, A.~Sozzetti\inst{3}
          }

   \institute{INAF -- Osservatorio Astronomico di Brera, Via E. Bianchi 46, 23807 Merate (LC), Italy
   \and
INAF -- Osservatorio Astrofisico di Arcetri, Largo E. Fermi 5, 50125 Firenze, Italy
\and
INAF -- Osservatorio Astrofisico di Torino, Via Osservatorio 20, 10025, Pino Torinese, Italy
\and
Dipartimento di Fisica, Universit\`{a} degli Studi di Torino, via Pietro Giuria 1, I-10125 Torino, Italy
\and
Space Research Institute, Austrian Academy of Sciences, Schmiedlstrasse 6, A-8042 Graz, Austria
\and
INAF -- Osservatorio Astrofisico di Catania, Via S. Sofia 78, 95123, Catania, Italy
\and
INAF -- Osservatorio Astronomico di Padova, Vicolo dell'Osservatorio 5, 35122, Padova, Italy
\and
Thüringer Landessternwarte Tautenburg, Sternwarte 5, 07778, Tautenburg, Germany
\and
INAF -- Osservatorio Astronomico di Palermo, Piazza del Parlamento, 1, 90134, Palermo, Italy
\and
Fundaci{\'o}n Galileo Galilei - INAF, Rambla Jos{\'e} Ana Fernandez P{\'e}rez 7, 38712 Bre$\tilde{\rm n}$a Baja, TF - Spain
\and
INAF -- Osservatorio Astronomico di Trieste, via Tiepolo 11, 34143 Trieste, Italy
\and
Department of Physics, University of Warwick, Coventry CV4 7AL, UK
\and
Centre for Exoplanets and Habitability, University of Warwick, Gibbet Hill Road, Coventry CV4 7AL, UK
\and
Astronomy Department, 96 Foss Hill Drive, Van Vleck Observatory 101, Wesleyan University, Middletown, CT 06459, US
\and
INAF -- Osservatorio Astronomico di Capodimonte, Salita Moiariello 16, 80131, Napoli, Italy
\and 
Dipartimento di Fisica e Chimica Emilio Segr{\'e} - Università di Palermo, Piazza del Parlamento, 1, 90134, Palermo, Italy
\and
Centro de Astrobiolog{\'i}a (CSIC-INTA), Carretera de Ajalvir km 4 - 28850 Torrej{\'o}n de Ardoz, Madrid, Spain
\and
Department of Physics, University of Rome Tor Vergata, Via della Ricerca Scientifica 1, I-00133 Rome, Italy
\and
Max Planck Institute for Astronomy, K\"{o}nigstuhl 17, D-69117, Heidelberg, Germany 
\and
INAF -- Osservatorio di Cagliari, via della Scienza 5, I-09047 Selargius, CA, Italy
\and
Dip. di Fisica e Astronomia Galileo Galilei -- Universit$\grave{\rm a}$ di Padova, Vicolo dell'Osservatorio 2, 35122, Padova, Italy
\and
Anton Pannekoek Institute for Astronomy, University of Amsterdam, Science Park 904, 1098 XH Amsterdam, The Netherlands
\and
INAF -- Osservatorio Astronomico di Roma, Via Frascati 33, 00078 Monte Porzio Catone, Italy
\and
Fondazione GAL Hassin, Via della Fontana Mitri, 90010 Isnello (PA), Italy
             }
             \offprints{F.~Borsa\\ \email{francesco.borsa@inaf.it}}

   \date{Received ; accepted }

% \abstract{}{}{}{}{} 
% 5 {} token are mandatory
 
  \abstract
   {}
   {In the framework of the GAPS project, we observed the planet-hosting star KELT-9 (A-type star, $v\sin{i}\sim$110\kms) with the HARPS-N spectrograph at the Telescopio Nazionale Galileo. In this work we analyse the spectra and the extracted radial velocities to constrain the physical parameters of the system and to detect the planetary atmosphere of KELT-9b.}
   {We extracted  the mean stellar line profiles from the high-resolution optical spectra via an analysis based on the least-squares deconvolution technique. Then we computed the stellar radial velocities with a method optimised for fast rotators by fitting the mean stellar line profile with a purely rotational profile instead of using a Gaussian function.}
   {The new spectra and analysis led us to update the orbital and physical parameters of the system, improving in particular the value of the planetary mass to $M_{\rm p} = 2.88 \pm 0.35\,\rm M_{Jup}$.
   We discovered an anomalous in-transit radial velocity deviation from the theoretical Rossiter-McLaughlin effect solution, calculated from the projected spin-orbit angle $\lambda=-85.78\pm0.46$ degrees measured with Doppler tomography. We prove that this deviation is caused by the planetary atmosphere of KELT-9b, thus 
   we call this effect Atmospheric Rossiter-McLaughlin effect. By analysing the magnitude of the radial velocity anomaly, we obtained information on the extension of the planetary atmosphere as weighted by the model used to retrieve the stellar mean line profiles, which is up to $1.22 \pm 0.02$ $\rm R_{\rm p}$.}
   {The Atmospheric Rossiter-McLaughlin effect will be observable for other exoplanets whose atmosphere has non-negligible correlation with the stellar mask used to retrieve the radial velocities, in particular ultra-hot Jupiters with iron in their atmospheres. The duration and amplitude of the effect will depend not only on the extension of the atmosphere, but also on the in-transit planetary radial velocities and on the projected rotational velocity of the parent star.}
   
   \maketitle
%
%________________________________________________________________
\section{Introduction\label{sec:intro}}

During the last two decades we have  witnessed  a huge expansion in exoplanet science.
By analysing demographic trends in exoplanet populations, many surprises and questions have emerged. This means that a larger observed sample (and thus more extensive surveys) is crucial in order to have a comprehensive view of exoplanetst \citep[e.g.][]{2015ARA&A..53..409W}.
We are finding planetary architectures that are completely different from  theoretical predictions, including extreme planetary systems. This diversity of exoplanetary systems reflects  an incredible heterogeneity of exoplanetary atmospheres \citep[e.g.][]{2016Natur.529...59S}. 
Thanks to  technological and methodological progress, we are increasing our ability to probe the atmospheric composition of other planets with the help of many different techniques \citep[e.g.][]{2017JGRE..122...53D}.
In this context, it is necessary to have small uncertainties on the parameters of the whole star--planet system to deeply characterise the physical and chemical processes of exoplanetary climates and interiors \citep{2016SSRv..205..285M}.

\object{KELT-9} \citep{2017Natur.546..514G} is one of the most striking examples of extreme systems: orbiting an A-type star with an effective temperature of $\sim$10000 K with a period of 1.48 days, the ultra-hot Jupiter \object{KELT-9b} reaches dayside equilibrium temperatures comparable to those of late K-type stars. 
At these high temperatures the atmospheric conditions are extreme \citep{2018ApJ...863..183K,2018ApJ...866...27L}, and it has been possible to recover some properties that 
were unexpected for planets. 

\citet{2018NatAs...2..714Y} detected hydrogen Balmer (H$\alpha$) absorption in the planetary atmosphere, extending up to the typical location of the thermosphere, indicating the presence of escaping excited hydrogen. 
They derived a planetary mass-loss rate of about 10$^{12}$ g\,s$^{-1}$ from the hydrogen atmospheric density extracted from the data, hence ignoring the effect of the heating provided by the stellar high-energy emission (i.e. in the extreme ultraviolet, XUV), which led them to suggest that the mass-loss rate may be even higher. However, as shown by \citet{2018ApJ...868L..30F}, this estimate is expected to be correct because early A-type stars, such as KELT-9, do not possess  significant XUV emission, which would further increase mass loss.
This can also explain why the atmosphere of KELT-9b has not completely escaped without requiring a short-lived evolutionary phase \citep{2018ApJ...868L..30F}.
\citet{2018Natur.560..453H} have found evidence of neutral and singly ionised iron and singly ionised titanium in the planetary atmosphere, while \citet{2019AJ....157...69C} have published the first detection of magnesium in an exoplanet and the presence of a dynamic atmosphere.

\smallskip

In this work, we update the KELT-9 system orbital and physical parameters and show that the planetary atmosphere can be detected through residual line profiles and in-transit radial velocity (RV) measurements. We find an in-transit RV deviation ascribable not only to the classical Rossiter-McLaughlin (RML) effect, which is caused by the Doppler shadow of the planet on the stellar mean line profile, but also by another effect arising from the planetary atmosphere detected thanks to the match with the stellar mask used to recover the stellar mean line profile.

%__________________________________________________________________

\section{Data sample\label{sec:data_sample}}

In the framework of the GAPS project \citep{2013A&A...554A..28C}, we observed KELT-9 during four transits of KELT-9b and twice out of transit 
with the HARPS-N and GIANO-B high-resolution spectrographs, mounted at the Telescopio Nazionale Galileo. We used the GIARPS configuration \citep{giarps}, that allowed us to observe simultaneously with the two spectrographs, obtaining high-resolution spectra in the wavelength range $\sim$390-690 $\rm nm$ and $\sim$940-2420 $\rm nm$.
In this work, we analyse only the HARPS-N spectra. The GIANO-B spectra will be presented in a future work, aimed at an in-depth atmospheric characterisation of the planet.

The transit of KELT-9b lasts $\sim$4 hours. The first transit observations were taken with low-cadence sampling with long exposures (600 sec), while for the other three transits 
we used a higher cadence (300 sec exposures).
Fiber A of the spectrograph was centred on the target, and at the same time fiber B was monitoring the sky. A log of the observations is given in Table~\ref{tab:log}. 
During the observations of transits 1 and 2 (and during the out-of-transit pointings) there was a problem with the atmospheric dispersion correctors (ADC) of the telescope, which remained stuck in the same position for the  duration of each observing block and was updated only after re-pointing the object. This caused a significant loss of flux, in particular in the blue part of the spectrum.
We note that our final results are valid even when considering only transits 3 and 4, and thus are not affected by the ADC problems.

\begin{table}
\begin{center}
\caption{KELT-9b HARPS-N observations log.}
\label{tab:log}
\footnotesize
\begin{tabular}{ccccc}
 \hline\hline
 \noalign{\smallskip}
Transit number & Night & Exposure time & N$_{\rm obs}$ & S/N$_{\rm ave}$\\
 \noalign{\smallskip}
 \hline
\noalign{\smallskip}
1 & 10 Jun 2018  & 600 s &36 & 156\\
2 & 23 Jul 2018  & 300 s &68 & 104\\
3 & 01 Sep 2018  & 300 s &58 & 118\\
4 & 04 Sep 2018  & 300 s &64 & 102\\
out of transit & 08 Jul 2018 & 300 s & 58 & 78 \\
out of transit & 22 Jul 2018 & 300 s & 89 & 135 \\
\noalign{\smallskip}
 \hline
\end{tabular}
\end{center}
\end{table}

The data were reduced with the version 3.7 of the HARPS-N Data Reduction Software (DRS) pipeline using the YABI interface with custom parameters \citep[e.g.][]{borsa}, in particular enlarging the cross-correlation function (CCF) width because of the large $v\sin{i}$ of the star. However, due to the ADC problems (and possibly also to the spectral type of the star, which does not have a proper colour correction template), we could not trust the given CCFs, which suffered from a clearly imperfect continuum normalisation. We thus started our analysis from the reduced one-dimensional spectra after they had been properly normalised.
We collected a total of 373 spectra, 226 bracketing four transits and 147 at other orbital phases.
We excluded from our final transit RV analysis all the data with a signal-to-noise ratio (S/N) <70 (as measured in order 46) or taken at airmass>1.7, thus discarding 11 of the 226 exposures collected. Adopting the transit ephemerides from \citet{2017Natur.546..514G}, we analysed 148 in-transit and 67 out-of-transit exposures.

%__________________________________________________________________

\section{Radial velocity extraction\label{sec:rv}}

The extraction of RVs from echelle spectra is usually done with a Gaussian fit on the CCFs of the different orders, properly combined with a weighted average \citep{baranne,2002A&A...388..632P}. 
For fast rotators, this procedure leads to large uncertainties on the derived radial velocity value as the rotational broadening is too large 
to ensure sufficient  precision. 
As we could not exploit the CCFs from the DRS, we extracted the RVs with a custom-made routine.
We computed the stellar mean line profiles using the least-squares deconvolution (LSD) software \citep{1997MNRAS.291..658D}. This software performs a deconvolution via a least-squares analysis of the normalised spectra  with a theoretical line mask extracted from the Vienna Atomic Line Database  \citep[VALD; ][]{piskunov}.
 We used a stellar mask with T$_{\rm eff}$=10000 K, log$g$=4.0, solar metallicity.
We accurately re-normalised the spectra and converted them to the required format, working only on the wavelength regions $4415-4805\AA$, $4915-5285\AA$, $5365-5870\AA$, $6050-6265\AA$, and $6335-6450\AA$; that is, we  cut the blue orders where the S/N was very low due to the instrument efficiency (and the ADC problems), the Balmer lines, and the regions where most of the telluric lines are found.

To estimate the RVs, instead of using a Gaussian fit, we preferred to model the LSD lines with a rotational profile, using the formula in Eq.~\ref{eq:rot_prof} \citep{gray}:

\begin{equation}
f(x)=1-2a(1-\mu)\sqrt{1-\left(\frac{x-x_0}{x_l}\right)^{2}} + \frac{0.5\pi\mu\left[1-\left(\frac{x-x_0}{x_l}\right)^{2}\right]}{{\pi}x_l\left(1-\frac{\mu}{3}\right)}
\label{eq:rot_prof}
\end{equation}

Here $a$ is the depth of the profile, $x$ the Doppler velocity, $x_0$ the centre  (i.e. the RV value), $x_l$ the $v\sin{i}$ of the star, $\mu$ the linear limb darkening (LD) coefficient.
The theoretical linear LD coefficient, calculated using the code \texttt{LDTK} \citep[which generates custom LD coefficients using a library of \texttt{PHOENIX}-generated specific intensity spectra; ][]{hannu,husser} for the same wavelength range used for the RV extraction and with our stellar parameters (Sect.~\ref{sec:orbit}) is $\mu=0.47$. However, here we decided to adopt the value $\mu=0.6$ because it minimises the RV residuals in the orbital fit (Sect.~\ref{sec:orbit}). This also reflects  the ADC problems encountered (see Sect.~\ref{sec:data_sample}), which  strongly reduced the flux at blue wavelengths, and introduced a strong (and not quantifiable) bias in the flux distribution along the wavelength range.

\begin{figure}%[!ht]
\centering
\includegraphics[width=\linewidth]{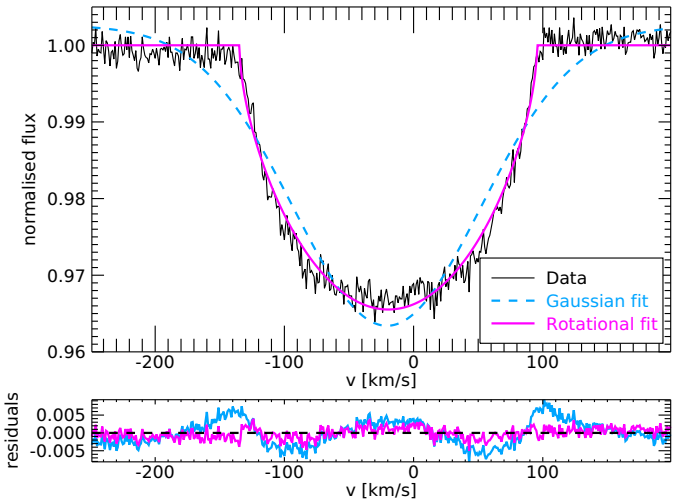}
\caption{Example of an out-of-transit stellar mean line profile of KELT-9. Also shown are the  Gaussian (cyan dashed line) and rotational profile (magenta line) fits. 
}
\label{fig:fit_lineprofile}
\end{figure}

Figure~\ref{fig:fit_lineprofile} shows an example of a mean line profile.
It is evident that the rotational profile is by far more suitable than a Gaussian function for fitting the line profile of fast rotators. A similar choice for extracting the RVs was also done by   \citet{2018arXiv180904897A} and \citet{2018AJ....155..100J}, among others.
The extracted RVs for our KELT-9 HARPS-N observations are listed in Table~\ref{tab:rv}.

We also derived the $v\sin{i}$ of the star from the rotational broadening fit ($x_l$ parameter in Eq.~\ref{eq:rot_prof}). Our result is $v\sin{i}$=111.8$\pm$1.0 \kms, taken as the average value calculated on the  out-of-transit mean line profiles with the relative standard deviation. 
We checked this value using the Fourier transform method on the line profiles \citep{1976oasp.book.....G,reiners}. The position of the first two zeros of the Fourier transform of a line profile is directly related to the $v\sin{i}$ when the rotational broadening is the largest broadening effect. This method yields a value of 110.0$\pm$3.0 \kms, which is consistent with our $v\sin{i}$ measurement.

\begin{table}
\begin{center}
\caption{HARPS-N RV observations of KELT-9. This table is available in its entirety online at the CDS.}
\label{tab:rv}
\footnotesize
\begin{tabular}{ccc}
 \hline\hline
 \noalign{\smallskip}
Time [BJD-2450000] & RV [\kms] & RV error [\kms]\\
 \noalign{\smallskip}
 \hline
\noalign{\smallskip}
  8280.46931 & -19.565 & 0.295 \\
  8280.47643 & -19.804 & 0.313 \\
  8280.48363 & -19.663 & 0.263 \\
  8280.49091 & -19.829 & 0.314 \\
 ... & ...  & ...\\
\noalign{\smallskip}
 \hline
\end{tabular}
\end{center}
\end{table}

  %************************************
%______________________________________________  
\section{Updated parameters of the planetary system\label{sec:orbit}}

We estimated the stellar atmospheric parameters by comparing the average of all collected HARPS-N spectra put in the stellar restframe with synthetic ones. In particular, we employed spectral synthesis with the {\sc synth3} code \citep{synth3} and the tools described by \citet{fossati2007} to measure the abundance of FeI and FeII from 21 and 30 lines, respectively, for values of the effective temperature (\Teff) ranging between 9000 and 10200\,K and surface gravity (\logg) ranging between 3.9 and 4.3. We then derived \Teff\ by imposing excitation equilibrium and \logg\ by imposing ionisation equilibrium. We computed stellar atmosphere models using LLmodels \citep{llmodels}, which was developed specifically to study the atmospheres of A-type stars. Prior to proceeding with the line fits, we measured the stellar \vsini \ by fitting the profiles of several weakly blended or unblended lines obtaining a value of 108$\pm$2\,\kms, in agreement with our other estimates. The broad line profiles did not allow a  direct measurement of the microturbulence velocity (\vmic), hence we assumed it to be equal to 1\,\kms, which has been shown to be an adequate value for early A-type stars \citep[e.g.][]{fossati2009}. We finally obtained \Teff\,=\,9600$\pm$400\,K and \logg\,=\,4.1$\pm$0.3 $\rm cm \ s^{-2}$. The wings of hydrogen Balmer lines are best suited to derive \Teff\ and \logg\ of A-type stars, particularly above 8500\,K \citep[e.g.][]{fossati2009}, but low-frequency waves in the spectra did not enable us to perform an acceptable normalisation of these broad features. With the derived atmospheric parameters, we measured an iron abundance [Fe/H]=+0.14$\pm$0.3\,dex.

\begin{figure}%[h!]
\centering
\includegraphics[width=\linewidth]{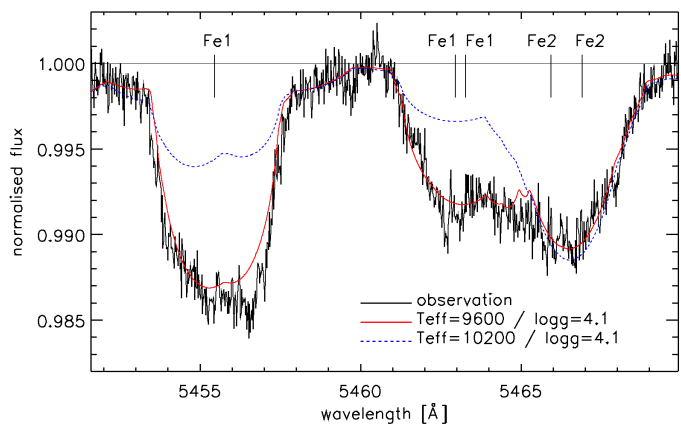}
\caption{Comparison of the average HARPS-N spectrum (black solid line), a synthetic spectrum obtained with the parameters derived from our analysis (red solid line), and a synthetic spectrum computed with the parameters of \citet[][blue dashed line]{2017Natur.546..514G}.}
\label{fig.comparison}
\end{figure}

The value of \Teff\ we obtained from the HARPS-N spectra is about 600\,K lower than that found by \citet{2017Natur.546..514G}, but both values actually agree within 1$\sigma$ given the large uncertainties of $\sim 400$~K. Figure~\ref{fig.comparison} shows a small portion of the average HARPS-N spectrum in comparison to synthetic spectra computed with the parameters given by \citet{2017Natur.546..514G} and derived from our analysis: our parameters better fit the observed spectrum. 

%--------------------------------------------------------------------
%--------------------------------------------------------------------

Since the KELT-9 parallax provided by the Gaia second Data Release \citep[DR2,][]{2018A&A...616A...1G}, 
that is $\pi=4.862\pm0.037$~mas, is significantly more precise than that used in \citet{2017Natur.546..514G}, we also re-determined the stellar parameters by fitting the spectral energy distribution (SED) and using the MIST evolutionary tracks \citep{2016ApJS..222....8D} 
through the ExofastV2 code \citep{2017ascl.soft10003E}. We first corrected $\pi$ for the systematic offset of $-82 \pm 33$~$\rm \mu as$ 
in the Gaia DR2 parallaxes reported in \citet{2018ApJ...862...61S}, then we used the same magnitudes as provided in Table~5 of \citet{2017Natur.546..514G} and modified the ExofastV2 code to include the stellar fluxes from the Ultraviolet Sky Survey Telescope. 
The stellar parameters are simultaneously constrained by the SED and the MIST isochrones, imposing Gaussian priors on \Teff\ and [Fe/H] from our updated stellar atmospheric parameters. The SED primarily constrains the stellar radius and temperature, and a penalty for straying from the MIST evolutionary tracks ensures that the resulting star is physical.
The derived stellar parameters are shown in Table~\ref{starplanet_param_table} and the mass, radius, and age 
agree with those of \citet{2017Natur.546..514G} within $1\sigma$.

\begin{figure}[t!]
\centering
\includegraphics[width=\linewidth, angle=90]{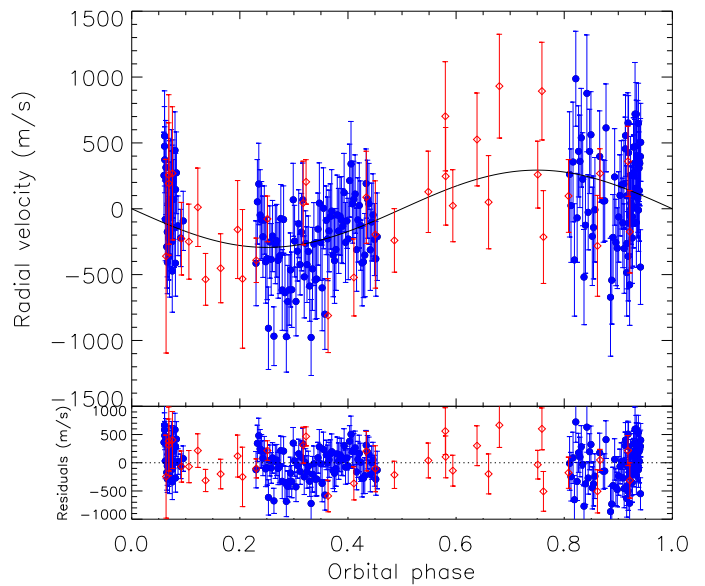}
\vspace{0.3cm}
\caption{
HARPS-N (blue circles) and TRES (red diamonds) out-of-transit radial velocities of KELT-9 as a function of the orbital phase along with the best Keplerian model (black solid line). The mid-transit corresponds to phases equal to zero or one.}
\label{fig_RV}
\end{figure}

We modelled the TRES RVs from literature \citep{2017Natur.546..514G} and our HARPS-N RVs with a Keplerian circular orbit in a Bayesian framework 
by employing a differential evolution Markov chain Monte Carlo (DE-MCMC) technique \citep{TerBraak2006, Eastmanetal2013}.
The circular orbit is justified by i) the strong constraint on the eccentricity from the observations of the secondary eclipse, that is $e < 0.015$ \citep[][Extended Data Table 3]{2017Natur.546..514G}, and ii) the expected circularisation of the orbit for star--planet tidal interactions, given that the circularisation timescale of $\sim$2-5 Myr for a modified tidal quality factor  $Q'_{\rm p}$$\sim10^6$ is considerably shorter than the system age ($\sim$450 Myr).
We fitted seven free parameters: the mid-transit time $T_{\rm c}$, the orbital period $P$, 
the RV semi-amplitude $K$, and two RV zero points ($V_{\rm r, HN}$ and $V_{\rm r, TRES}$) 
and jitter terms ($s_{\rm j,HN}$ and $s_{\rm j,TRES}$) for the two spectrographs HARPS-N and 
TRES, respectively. We imposed Gaussian priors on $T_{\rm c}$ and $P$ from the orbital ephemeris in \citet{2017Natur.546..514G}
and uninformative priors on all the other parameters. 
We ran 14 DE-MCMC chains by adopting the same likelihood function and prescriptions as in \citet{Bonomoetal2017}. 
The medians and the 15.86\% and 84.14\% quantiles of the posterior distributions were taken as the best
values and $1\sigma$ uncertainties, and are listed in Table~\ref{starplanet_param_table}. 
The Keplerian best fit is shown in Fig.~\ref{fig_RV}. 
Unlike \citet{2018NatAs...2..714Y} who estimated the planetary RVs from the planetary atmospheric track (using H$\alpha$), we did not because in our data we detected possible hints of atmospheric dynamics that could cause a deviation from the planetary Keplerian motion; this effect will be analysed in another paper.

Finally, we combined the posterior distributions of the stellar, orbital, and transit parameters (the last  from \citealt{2017Natur.546..514G}) to redetermine the planetary parameters (see Table~\ref{starplanet_param_table}). 
In summary, we find a radius $R_{\rm p}=1.936 \pm 0.047\,\rm R_{Jup}$, 
a mass $M_{\rm p} = 2.88 \pm 0.35\,\rm M_{Jup}$, 
and thus a density $\rho_{\rm p}=0.491^{+0.072}_{-0.066}$\,g\,cm$^{-3}$ for KELT-9b. 
These planetary parameters are certainly consistent with those that were previously determined by \citet{2017Natur.546..514G}, 
but are more precise. In particular, the significance of our 
planetary mass measurement increased from 3.4$\sigma$ to $8.2\sigma$, 
and that of the planetary density from 3.5$\sigma$ to $7.4\sigma$, which is a remarkable improvement 
given the large RV uncertainties due to the high $v\sin{i}$ of the star.

%*************************************************************

\begin{table*}[t]
\centering
\caption{Properties of the KELT-9 planetary system.}            
\begin{minipage}[t]{13.0cm} 
\renewcommand{\footnoterule}{}                          
\begin{tabular}{l l l}        
\hline\hline                 
\emph{Stellar IDs} &  & \\
\hline
HD 195689 & & \\
TYC 3157-638-1 & & \\
2MASS J20312634+3956196 & & \\
& \\
\hline
\emph{Stellar parameters}  &  & Source \\
\hline
Star mass $M_\star$ [\Msun] &  $2.32 \pm 0.16$ & This work \\
Star radius $R_\star$ [\Rsun] & $2.418 \pm 0.058$ & This work  \\
Stellar luminosity $L_\star$ [${L_\odot}$] & $38.9^{+4.3}_{-3.2}$ & This work \\
Stellar density $\rho_{*}$ [$ \rm g\;cm^{-3}$] & $0.231 \pm 0.025$ & This work \\
Effective temperature $T_{\rm{eff}}$[K] & $9600 \pm 400$ & This work~$^a$ \\
Derived effective temperature $T_{\rm{eff}}$[K] & $9270^{+240}_{-180}$ & This work~$^b$ \\
Surface gravity log\,$g$ [cgs] &  $4.1 \pm 0.3$ & This work~$^a$ \\
Derived surface gravity log\,$g$ [cgs] &  $4.037^{+0.038}_{-0.040}$ & This work~$^b$ \\
Metallicity $[\rm{Fe/H}]$ [dex] & $0.14 \pm 0.30$ & This work~$^a$ \\
Derived metallicity $[\rm{Fe/H}]$ [dex] & $0.07^{+0.20}_{-0.23}$ & This work~$^b$ \\
Age $t$ [Gyr]  & $0.45^{+0.14}_{-0.13}$  & This work \\
HARPS-N systemic velocity  $V_{\rm r, HN}$ [\kms] & $-19.819 \pm 0.024$ & This work \\
TRES radial-velocity zero point  $V_{\rm r, TRES}$ [\kms] & $ 0.391\pm 0.057$ & This work \\
HARPS-N radial-velocity jitter  $s_{\rm j,HN}$ [\kms] & $0.075 \pm 0.048$ & This work \\
TRES radial-velocity jitter  $s_{\rm j,TRES}$ [\kms] & $0.138 \pm 0.074$ & This work \\
Gaia DR2 parallax $\pi$ [mas] & $4.944\pm0.050$ & This work \\  
Interstellar extinction $A_V$ & $0.106^{+0.077}_{-0.064}$ & This work \\
Stellar distance $d$ [pc] & $202.3\pm2.0$ & This work \\
& \\
\hline
\emph{Transit and orbital parameters}  &  & Source \\
\hline
Orbital period $P$ [d] & $1.4811235 \pm 0.0000011$ & \citet{2017Natur.546..514G} \\
Transit epoch $T_{ \rm 0} [\rm BJD_{TDB}-2450000$] & $7095.68572 \pm 0.00014$ & \citet{2017Natur.546..514G} \\
Radius ratio $R_{\rm p}/R_{*}$ & $0.08228 \pm 0.00043$ & \citet{2017Natur.546..514G} \\
Inclination $i$ [deg] & $86.79 \pm 0.25$  & \citet{2017Natur.546..514G} \\
Orbital eccentricity $e$  &  $0~\rm (fixed) $ & \citet{2017Natur.546..514G} \\
Radial velocity semi-amplitude $K$ [\ms] & $293 \pm 32 $ & This work \\
& \\
\hline
\emph{Planetary parameters} & & Source \\
\hline
Planet mass $M_{\rm p} ~[\Mjup]$  &  $2.88\pm0.35$ & This work \\
Planet radius $R_{\rm p} ~[\Rjup]$  &  $1.936\pm0.047$ & This work  \\
Planet density $\rho_{\rm p}$ [$\rm g\;cm^{-3}$] &    $0.491_{-0.066}^{+0.072}$ & This work \\
Planet surface gravity log\,$g_{\rm p }$ [cgs] &   $3.279_{-0.058}^{+0.053}$ & This work  \\
Orbital semi-major axis $a$ [AU] & $0.03368\pm0.00078$ & This work \\
Equilibrium temperature $T_{\rm eq}$ [K]~$^c$  & $3921_{-174}^{+182}$ & This work \\
\hline       
\hline
\vspace{-0.3cm}
\footnotetext[1]{\scriptsize from the analysis of HARPS-N spectra.} \\
\footnotetext[2]{\scriptsize from SED fitting and MIST evolutionary tracks by using the ExofastV2 tool and imposing Gaussian priors on $T_{\rm{eff}}$ and $[\rm{Fe/H}]$ from the spectral analysis.} \\
\footnotetext[3]{\scriptsize black-body equilibrium temperature assuming a null Bond albedo and uniform heat redistribution to the night-side.} \\
\end{tabular}
\end{minipage}
\label{starplanet_param_table}  
\end{table*}

%______________________________________________________________
\section{In-transit radial velocities: the Atmospheric Rossiter-McLaughlin effect\label{sec:atmo_rml}}

\begin{figure*}%[!ht]
\centering
\includegraphics[width=\linewidth]{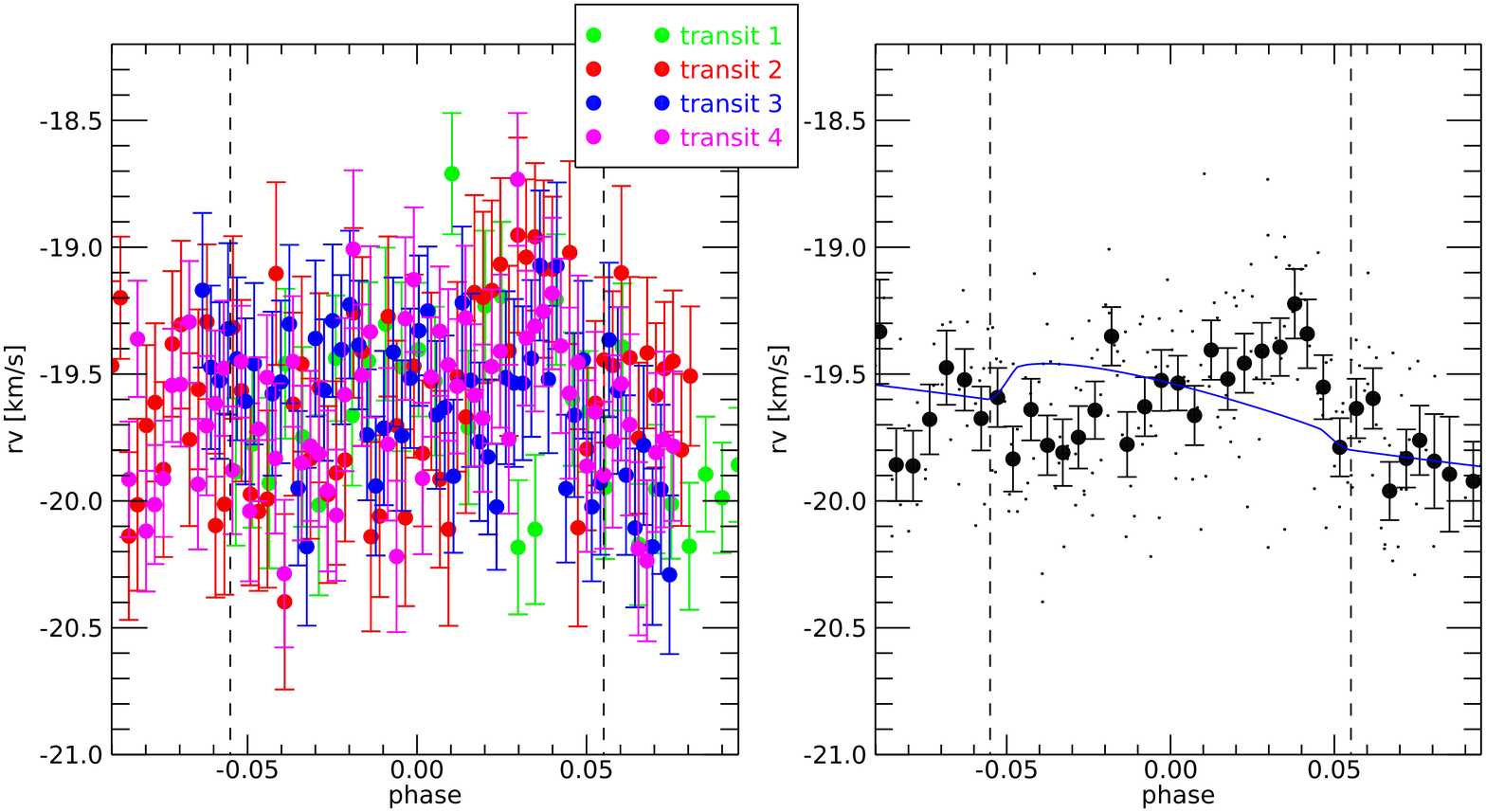}
\caption{{\it (Left panel)} Phase-folded RVs of the four HARPS-N transits of KELT-9b. Different colours refer to different transits. {\it (Right panel)} Mean RVs of the four transits (filled circles), together with single RVs (small dots). The theoretical RV solution (blue line) computed using Keplerian motion and the classical RML effect alone  clearly does not match the observations. The vertical dashed lines show the ingress and egress of transit.}
\label{fig:rml_doppio}
\end{figure*}

For each transit sequence we subtracted the mean in-transit RV from each RV
to avoid any possible offset caused by instabilities and long-term trends in the orbital solution, and then averaged the RVs of the four transits in bins of 0.005 in phase.
The RV time series of each single transit and their average are shown in Fig.~\ref{fig:rml_doppio}.

\begin{figure}%[!ht]
\centering
\includegraphics[width=\linewidth]{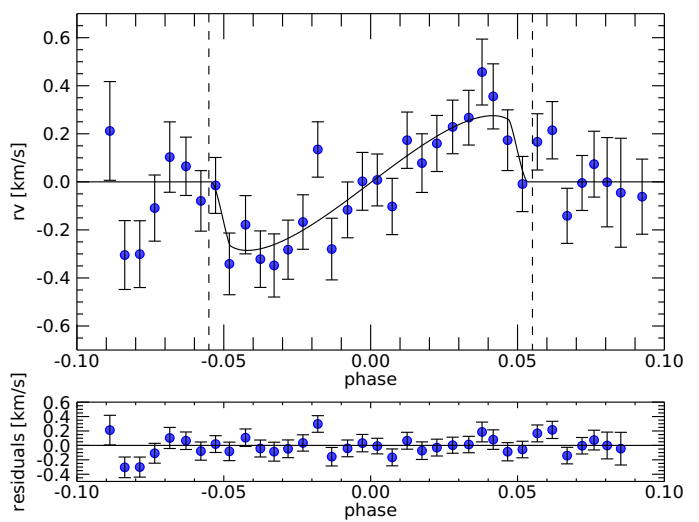}
\caption{Atmospheric RML effect of KELT-9b. The vertical dashed lines show the ingress and egress of transit.} 
\label{fig:rml_atmo}
\end{figure}

\begin{figure}%[!ht]
\centering
\includegraphics[width=\linewidth]{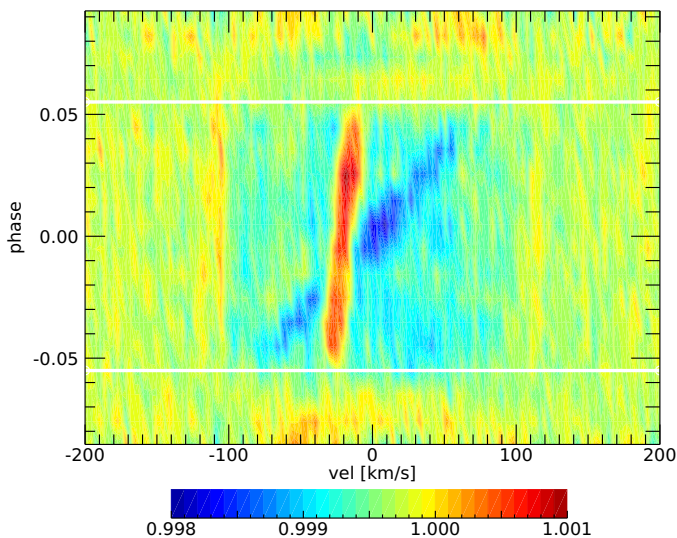}
\caption{Mean line profile residuals tomography of the four KELT-9 transits observed, centred in the stellar restframe. The horizontal white lines show the ingress and egress of transit. The doppler shadow of the planet (red) and the planetary atmosphere track (blue) are both evident. The  depression during the transit (light
blue) is caused by the photometric decrease of flux.}
\label{fig:tomography}
\end{figure}

For KELT-9b, by using the technique of line-profile tomography \citep[e.g.][]{2010MNRAS.403..151C} the projected orbital plane inclination with respect to the stellar rotational axis was found to be $\lambda=-84.8\pm1.4$ degrees \citep{2017Natur.546..514G}.
By applying their same Doppler shadow model to our data, we updated this value to $\lambda=-85.78\pm0.46$ degrees.
Combining this information with our $v\sin{i}$ value (Sect.~\ref{sec:rv}) and R$_{\rm p}$/R$_{\rm s}$=0.08228 \citep{2017Natur.546..514G}, we calculated the expected RML effect RV curve.
This theoretical curve, summed with the Keplerian orbital solution of Sect.~\ref{sec:orbit},  clearly does not match the in-transit RV time series (Fig.~\ref{fig:rml_doppio}, right panel).
Non-convective A-type stars like KELT-9 do not have spots. 
We thus investigated the possible reasons of the discrepancy between the observed RVs and the theoretical solution by looking at their differences (Fig.~\ref{fig:rml_atmo}) and by analysing the tomography of the mean line profiles.
This was created by dividing for each transit observed all the mean line profiles by a master out-of-transit mean line profile. All  four transits were then combined to obtain Figure~\ref{fig:tomography}.

The resulting RV residuals from the theoretical solution (Fig.~\ref{fig:rml_atmo}) clearly show another RML-like shape. This is actually not surprising as  in the tomography of the mean line profile residuals (Fig.~\ref{fig:tomography}) the planetary atmospheric track is evident, as it is  in \citet{2017Natur.546..514G} and \citet{2018Natur.560..453H,2019arXiv190502096H} as well. We could thus expect this signal, given that the planetary radial velocities are within the width of the stellar mean line profile for the whole transit duration.

We performed an RML fit to these residuals, using the IDL routine \textsc{MPFIT} and the model of \citet{ohta} as implemented in EXOFAST \citep{Eastmanetal2013}, fixing the $v\sin{i}$ and the linear limb darkening to the values we estimated in Sect.~\ref{sec:rv}. 
We also decided to keep the inclination angle $\lambda$ fixed at the value of -180 degrees. 
This choice reflects two facts. First the atmospheric planetary signal, that causes the deviation of the global in-transit stellar RVs, moves at a velocity 
corresponding to that of the planetary RV. Second, contrarily to what happens for the Doppler shadow which is associated with a relative flux increase (a positive `bump' moving along the mean line profile), it represents a further absorption.
%This choice reflects the fact that the atmospheric planetary signal, that causes the deviation of the global in-transit stellar RVs, moves at a velocity 
%corresponding to that of the planetary RV and, contrarily to what happens for the Doppler shadow, which is associated with a relative flux increase (a positive `bump' moving along the mean line profile) represents a further absorption.\LEt{ This sentence is very long and convoluted. Could it perhaps be broken into two parts? }
The planetary signal will thus cross the stellar mean line profile from negative to positive velocities, like in the case of the Doppler shadow of an aligned planet, but since it is giving a further absorption and not a reduction of it, the measured RV shift will have opposite sign. The effect on the RVs will thus be like that of a planet transiting with a projected obliquity of -180 degrees. We note that any possible deviation from -180 degrees could be a hint of eccentricity of the planetary orbit. Our choice to impose the value of  -180 degrees  is thus also consistent with the circular orbital solution adopted.
Leaving $\lambda$ as a free parameter in the fit, however, does not change its full compatibility with the imposed -180 degrees value ($\lambda$=-175$\pm$20 degrees). 
The fit of this {Atmospheric RML} gives a value of R$_{\rm p}$/R$_{\rm s}$=0.058$\pm$0.004. We interpret this as the R$_{\rm atmo}$/R$_{\rm s}$ of the atmospheric layer that matches the stellar model used to extract the mean line profiles.

%\subsection{Chromatic RVs\label{sec:atmo_chromatic}}

One way to detect the atmosphere of exoplanets using chromatic RVs was proposed by \citet{2004MNRAS.353L...1S}, who analysed the amplitude of the RML effect at different wavelengths to retrieve a transmission spectrum. If the planetary apparent radius varies with wavelength, this will result in a different amplitude of the RML effect. \citet{digloria} successfully applied  the technique to high-resolution spectroscopic observations of HD~189733b, finding evidence of Rayleigh scattering in its atmosphere.

We made an attempt to see if any chromatic effect caused by the planetary atmosphere could be detected in our RV measurements of KELT-9.
We split each spectrum into two parts, setting 5500 $\AA$ as the dividing value, and performed  the same analysis again, as described in Sect.~\ref{sec:rv}.
No significant difference is found with the R$_{\rm p}$/R$_{\rm s}$ from the Atmospheric RML fit. This is mainly due to the low S/N found in the stellar mean line profile of the redder half of the spectra, caused by the low number of stellar lines for such a spectral type.

As a second approach, we created stellar masks using only Ti, FeI, and FeII lines, respectively. The only case in which we reached sufficient RV precision to detect in-transit deviations from the Keplerian motion is with the FeII mask, for which the atmospheric RML effect results are fully compatible with those obtained using the complete A0 stellar mask. This is another confirmation of the strong presence of FeII in the atmosphere of KELT-9b, previously detected at >20$\sigma$ significance \citep{2018Natur.560..453H}.

%______________________________________________________________
\section{Discussion\label{sec:discussion}}

The atmosphere of KELT-9b, with a dayside equilibrium temperature $T_{\rm eq}\sim$4000 K, has been proven to host metals in a gaseous and excited state \citep{2018Natur.560..453H}.
As a result, it significantly correlates with the stellar model that we used to retrieve the stellar mean line profile. 
This leaves an imprint in the stellar mean line profile during each transit at the corresponding radial velocity of the planet and, as a consequence, affects the values of the retrieved radial velocities of the star.

We do not find any evidence of significant asymmetry in the atmospheric RML shape.
It is  interesting to note, however,  that the phase where the residual RVs show the largest discrepancy from the fit (Fig.~\ref{fig:rml_atmo}) is at $\sim-0.02$, 
which is the same phase where the two tracks of the 
Doppler shadow and planetary atmosphere intersect (Fig.~\ref{fig:tomography}). This deviation appears to happen for all the analysed transits (Fig.~\ref{fig:rml_doppio}, left panel), so it looks unlikely that it is just due to outliers. 
The deviation could be due to the partial cancellation that the Doppler shadow and the atmospheric track give to each other.
Another possibility is that it is caused by the  symmetric line profile adopted for fitting the LSD profiles, and thus for measuring the RVs
\citep{2005ApJ...631.1215W,2009A&A...506..377T}, and this effect is enhanced in the overlap of different asymmetries.

We hypothesise that the anomalous signal intercepted in our analysis of the stellar mean line profile is given only by the planetary atmosphere. Thus, the 
$R_{\rm atmo}$/R$_{\rm s}$ of the atmospheric RML fit  refers to the atmosphere as if it were a disk, and we compare 
the area of this disk to the value of the planetary area extracted from photometric transits.
We take as reference R$_{\rm p}$/R$_{\rm s}$=0.08228$\pm$0.00043 \citep{2017Natur.546..514G}.
Comparing the two resulting areas, the atmospheric area  represents $\sim$51\% of the photometric planetary area. 
We do not have any hint on the atmospheric shape, but if 
we hypothesise the most simple case of an atmosphere as an annulus surrounding the planetary core, this could  translate to a 
$R_{\rm p+atmo}= 1.22 \pm 0.02$ $R_{\rm p}$.
Considering that \citet{2018NatAs...2..714Y} found a radius at the H$\alpha$ line centre $\sim$1.64 times the size of the planetary radius, this is another confirmation of an extended atmosphere and that also heavy atoms are present at rather high altitudes.

%______________________________________________________________

\section{Conclusions\label{sec:conclusions}}

We extracted  RVs of KELT-9 from HARPS-N spectra with an appropriate model to derive radial velocities for fast rotators, by using a rotational profile instead of a Gaussian to fit the stellar mean line profiles.
This method led to more precise results.
With the new set of spectra and RVs and by exploiting the Gaia DR2 we were able to update the orbital and physical parameters of the system. 

We evidenced an in-transit RV deviation from the Keplerian orbital solution, which is not caused by the RML effect alone. This further deviation, which we call Atmospheric RML, is caused by the atmosphere of the transiting planet.
The Atmospheric RML signal is obtained analogously to previous detections of the transmission spectra of exoplanets HD~209458~b \citep{2010Natur.465.1049S} and HD~189733~b \citep{2016ApJ...817..106B,2018A&A...615A..16B} with near-infrared high-resolution spectroscopy, with the important difference that in those studies the search for exoplanetary signals was performed with template spectra appropriate for the atmosphere of the planet, and the star was seen  as a contaminating RML effect \citep[e.g.][]{2016ApJ...817..106B}, whereas in this case the planet signal is picked up with a mask optimised for the parent star. This is once again due to the exceptionally high planet temperature, which creates conditions more like  a stellar photosphere  than a classic planet atmosphere.

To our knowledge, this is the first time that such an effect has been observed in RVs. By analysing in detail this anomaly, we were able to determine the mean atmospheric height (as weighted with the stellar mask) of KELT-9b.
We note that in particular cases of projected spin-orbit angle close to $\lambda=0$, this effect could not be seen even if  present. However, it could lead to an underestimation of R$_{\rm p}$/R$_{\rm s}$ and/or $v\sin{i}$  when fitting for the RML effect. In this case the planetary atmospheric track and the Doppler shadow can completely overlap if the excursion of the in-transit planetary radial velocities is $\sim 2\times$ \vsini.%\LEt{ 1st conditional, likely. or: could overlap if the velocities were (2nd conditional, hypothetical) }

We also note that the duration of the Atmospheric RML coincides with that of the transit only when the in-transit planetary radial velocities are within the stellar mean line profile, as is the case for KELT-9. For other cases, depending on the $v\sin{i}$  of the star and on the orbital parameters of the system, this effect will only be present in the central phase of the transit when the planetary radial velocities fall within the stellar mean line profile range. 
This could lead to strange in-transit RV variations.
The Atmospheric RML will be observable for other exoplanets whose atmospheres have a non-negligible correlation with the stellar mask used to retrieve the radial velocities, in particular ultra-hot Jupiters with iron in their atmospheres.

%______________________________________________________________

\begin{acknowledgements}
We thank the referee for the useful comments that helped improve the clarity of the paper.
We acknowledge the support by INAF/Frontiera through the `Progetti Premiali' funding scheme of the Italian Ministry of Education, University, and Research.
FB acknowledges financial support from INAF through the ASI-INAF contract 2015-019-R0.
DB acknowledges financial support from INAF and the Agenzia Spaziale Italiana (ASI grant n. 014-025-R.1.2015) for the 2016 PhD fellowship programme of INAF.
This work has made use of data from the European Space Agency (ESA) mission {\it Gaia} 
(\url{https://www.cosmos.esa.int/gaia}), processed by the {\it Gaia} 
Data Processing and Analysis Consortium (DPAC, 
\url{https://www.cosmos.esa.int/web/gaia/dpac/consortium}). Funding for 
the DPAC has been provided by national institutions, in particular the 
institutions participating in the {\it Gaia} Multilateral Agreement.
\end{acknowledgements}

%______________________________________________________________

\end{document}